\documentclass[twocolumn,showpacs,preprintnumbers,amsmath,amssymb,prl]{revtex4}
\usepackage{graphicx}
\usepackage{dcolumn}
\usepackage{bm}

\begin{document}
\title{High-Precision Optical Measurement of the 2S Hyperfine Interval in Atomic
Hydrogen}

\author{N. Kolachevsky}
\altaffiliation[Also at P.N. Lebedev Physics Institute, Moscow,
Russia] \
\author{M. Fischer}
\author{S.G. Karshenboim}
\altaffiliation[Also at D.I. Mendeleev Institute for Metrology,
St. Petersburg, Russia] \
\author{T.W. H\"{a}nsch}
\altaffiliation[Also at Ludwig-Maximilians-University, Munich,
Germany] \ \affiliation{Max-Planck-Institut f\"{u}r Quantenoptik,
Garching, Germany}

\date{\today}

\begin{abstract}
We have applied an optical method to the measurement of the $2S$
hyperfine interval in atomic hydrogen. The interval has been
measured by means of two-photon spectroscopy of the $1S-2S$
transition on a hydrogen atomic beam shielded from external
magnetic fields. The measured value of the $2S$ hyperfine interval
is equal to $177\ 556\ 860(15)$ Hz and represents the most precise
measurement of this interval to date. The theoretical evaluation
of the specific combination of $1S$ and $2S$ hyperfine intervals
$D_{21}$ is in moderately good agreement with the value for
$D_{21}$ deduced from our measurement. 
\end{abstract}

\maketitle

The frequency of the $2S$ hyperfine interval $f_{\rm HFS}({ 2S})$
has been measured twice during the last 50 years by driving the
magnetic-dipole radio-frequency transition in a hydrogen thermal
beam \cite{Kusch,Hessels}. The relative accuracy of these
measurements ($150-300$ ppb) exceeds the accuracy of the
theoretical prediction for the $2S$ hyperfine interval which is
restricted by an insufficient knowledge of the proton structure.
However, the specific combination of the $1S$ and $2S$ hyperfine
intervals
\begin{equation}
 D_{21} = 8 f_{\rm HFS}({2S}) - f_{\rm HFS}({1S})
\label{eq1}
\end{equation}
can be calculated with high precision due to significant
cancellations of nuclear structure effects (see \cite{Karsh1} and
references therein). As the $1S$  hyperfine splitting in hydrogen
$f_{\rm HFS}({1S})$, known experimentally to several parts in
$10^{13}$ (see e.g. \cite{Ramsey}), does not restrict the accuracy
of (\ref{eq1}), it is possible to compare the experimentally
measured $2S$ hyperfine interval with $f_{\rm HFS}({2S})$ deduced
from the theoretical $D_{21}$ value. The quantum-electrodynamics
theory (QED) for the state-dependent contribution to the $nS$
hyperfine splitting can thus be tested to the level up to
$\alpha^4$ and $\alpha^3 m_e/m_p$. This test is limited only by
the experimental uncertainty.

The recent theoretical value of $D_{21}^{\rm theor}$ is equal to
$48 \ 953(3)$ Hz and corresponds to a $2S$ hyperfine interval of
$f_{\rm HFS}^{\rm theor}({2S})=177\ 556\ 838.1(4)$ Hz
\cite{Karsh1}.
In 1956 Heberle, Reich, and Kusch measured $f_{\rm HFS}({2S})$ for
the first time \cite{Kusch}. Their result was equal to $177\ 556\
860(50)$ Hz which is in an agreement with $f_{\rm HFS}^{\rm
theor}({2S})$. In 2000 Rothery and Hessels \cite{Hessels} improved
the accuracy and obtained the value of $177\ 556\ 785(29)$ Hz. We
have performed a totally independent optical measurement of
$f_{\rm HFS}({2S})$. The result of our measurement is $177\ 556\
860(15)$ Hz which is up to now the most precise value for the $2S$
hyperfine interval in atomic hydrogen. Both recent results are in
a moderately good agreement (within $2 \sigma$) with the
theoretical value.

For the measurement of the $2S$ hyperfine interval in atomic
hydrogen we have applied $1S-2S$ two-photon spectroscopy to a cold
hydrogen atomic beam which is shielded from magnetic fields. Using
a high-finesse cavity as a frequency flywheel we deduce the $2S$
hyperfine interval as the frequency difference between two
extremely stable laser light fields which excite the respective
transitions between the different hyperfine sublevels of the $1S$
and $2S$ states in atomic hydrogen. The differential measurement
cancels some important systematic effects typical for two-photon
spectroscopy on atomic beams. Applying this optical method, we
achieve a level of accuracy which is nearly 2 times better than
the accuracy of the recent radio-frequency measurement
\cite{Hessels}. Along with the previous optical Lamb shift
measurement \cite{Hbook}, our present measurement demonstrates the
perspectives of precision optical methods in fields where
traditionally radio-frequency techniques have been used.

The hydrogen spectrometer setup, described in detail elsewhere
\cite{Huber}, has been modified by magnetic compensation and
shielding systems and an optional differential pumping system
\cite{Fischer}. A dye laser operating near 486 nm is locked to an
ultra-stable reference cavity made from ULE by means of the
Pound-Drever-Hall lock. The drift of the cavity, suspended in a
vacuum chamber with a two-stage active temperature stabilization
system, is typically 0.5 Hz/s.

The frequency of the dye laser light is doubled in a
$\beta$-barium borate crystal, and the resulting UV radiation near
243 nm is coupled into a linear enhancement cavity inside a vacuum
chamber. Atomic hydrogen, produced in a radio-frequency discharge
at a pressure of around 1 mbar, flows through teflon tubes to a
copper nozzle cooled to 5 K with a helium flow-through cryostat.
Hydrogen atoms thermalize in inelastic collisions with the cold
walls of the nozzle. The atomic beam escapes from the nozzle
coaxially to the mode of the enhancement cavity. On their way
through the laser field some atoms are excited via Doppler-free
two-photon absorption from the ground state to the metastable $2S$
state. In the detection region, these atoms are quenched in a
small electric field and emit Lyman-${\alpha}$ photons which are
counted by a photomultiplier. Slow atoms are selected by time
resolved spectroscopy \cite{Huber} so that the second-order
Doppler shift and the time-of-flight broadening are reduced,
yielding typical linewidths around 2 kHz at 121 nm both for the
$1S(F=0) \rightarrow 2S(F=0)$ (singlet) and $1S(F=1) \rightarrow
2S(F=1)$ (triplet) transition lines.

A turbo pump evacuates the main volume of the vacuum system to $5
\times 10^{-5}$ mbar. The excitation region, separated from the
main volume by a non-magnetic metal housing, is differentially
pumped by a large cryopump. Two small holes in the front and back
walls of the housing allow the excitation light to enter and exit
this high vacuum zone and collimate the atomic beam. With hydrogen
atoms escaping from the cold nozzle, the pressure in the
excitation region is typically 3$\times 10^{-8} $ mbar. An
additional lockable opening in the housing allows to measure at a
worse pressure of $1.5 \times 10^{-7}$ mbar. By increasing the
temperature of the cryopump, it is also possible to work at even
higher pressures up to $5 \times 10^{-6}$ mbar. All parts adjacent
to the hydrogen beam are covered with graphite to reduce stray
electric fields in the excitation region which would quench the
hydrogen $2S$ population and shift the transition frequencies due
to the DC Stark effect.

 To reduce the magnetic field along the excitation region we
use a two-stage magnetic shielding setup together with external
compensation coils. We have measured the residual field inside the
first shielding stage made from 100 $\mu$m thin $\mu$-metal foil
which encloses the entire excitation region, the detector, and the
nozzle to be less than 20 mG. Inside this shielding, 1 mm thick
$\mu$-metal tubes located along the enhancement cavity axis cover
about 90\% of the whole excitation path of the hydrogen atoms. The
evaluated averaged shielding factor of the second shielding stage
is more than 20.

An external magnetic field shifts the magnetic sublevels of the
hydrogen $1S_{1/2}$ and $2S_{1/2}$ states according to the
Breit-Rabi equation \cite{Bethe}. For two-photon processes,
allowed transitions obey the selection rules $\Delta F=0$ and
$\Delta m_F=0$. In our experiment, we excite two-photon
transitions from different magnetic sublevels of the hydrogen
ground state to corresponding sublevels of the $2S$ state. In
small magnetic fields, when the triplet splitting vanishes, the 2S
hyperfine interval is given by
\begin{equation}
f_{\rm HFS}({2S}) = f_{\rm HFS}({1S})+f(1,\pm 1 {\ \rm or\ }
0)-f(0,0),  \label{eq2}
\end{equation}
where the symbol $f(F,m_F)$ denotes the transition frequency
between sublevels with quantum numbers $(n=1,F,m_F)$ and
$(n=2,F,m_F)$ at 121 nm. A magnetic field $H$ shifts $f_{\rm
HFS}({2S})$ approximately as $10 \times H^2$ kHz/$ \rm G^2$.

The dye laser is locked to a TEM$_{00}$ mode of the reference
cavity. Its frequency can be changed by means of a double-passed
broadband AOM  placed between the laser and the cavity. The
frequency shift corresponding to 121 nm is 8 times higher in
absolute value than the frequency shift of the synthesizer driving
the AOM. The factor 8 arises due to the double-passed AOM, the
optical frequency doubling, and the two-photon excitation of the
hydrogen atoms. The intensity of the light used to lock the laser
to the reference cavity is stabilized. All synthesizers providing
the radio frequencies in our experiment are locked to the 10 MHz
signal of a commercial HP5071A cesium frequency standard
(specified Allan standard deviation $5 \times 10^{-12}$ within one
second). The standard introduces a negligible error to the
measured value.

During 16 days of measurements we have recorded about 2000
hydrogen spectra for the triplet and singlet transitions. A single
spectrum consists of about 30 data points, each measured for 1
second. For our fitting procedure we have chosen the spectra
recorded at a delay time of $810 \ \mu$s (time between the
blocking of the excitation light and the start of photon
counting), for which the spectrum asymmetry is considerably
reduced. A typical count rate for the triplet transition in its
maximum is 350, while the averaged ratio between triplet and
singlet count rates is $3.25 \pm 0.03$. We ascribe the deviance of
this value from 3 to different recombination rates in the nozzle
for hydrogen atoms in the singlet and triplet ground states.

One measurement run consists of $2-6$ hydrogen spectra recorded
one after another within approximately 5 minutes. After each run,
we change the frequency of the laser light to excite the other
transition. During a measurement day, we have switched about 50
times between the triplet and singlet transitions. The intensity
of the excitation light is monitored after the enhancement cavity
and has been kept as constant as possible during the whole day of
measurement.

\begin{figure}
\begin{center}
\includegraphics [width=60 mm]{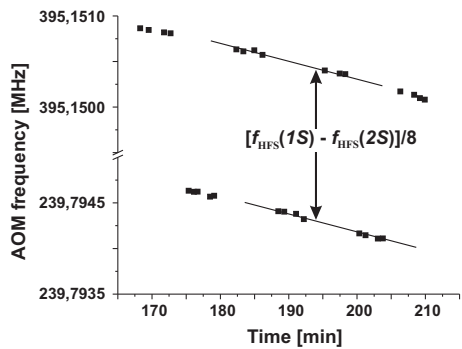}
\caption{AOM frequencies corresponding to the singlet (circles)
and triplet (squares) transitions. The frequency dependence of the
data is due to the cavity drift. The linear fit with the same
slope of four data runs is shown.}\label{singtrip}
\end{center}
\end{figure}

As the laser is locked to the same mode of the cavity, the cavity
drift is the same for both singlet and triplet transitions. To
determine the drift, we have fitted each hydrogen spectrum with a
Lorentzian function in the time and frequency domains. A part of a
day drift data set is shown on Fig.\ref{singtrip}. To reduce the
effect of a nonlinear cavity drift, the drift data are fitted
stepwise within the short time periods (about 20 minutes each)
covering 2 singlet and 2 triplet runs. During this time period the
drift can be well approximated linearly, and we have fitted the
data by means of linear regression with the same slope. The fit
procedure has been performed analytically delivering the slope
(cavity drift), offset frequency, and corresponding errors.
According to the hydrogen level scheme and the measurement
technique, the offset frequency is equal to $\left[ f_{\rm
HFS}({1S})-f_{\rm HFS}({ 2S})\right]/8 $. From this, the $2S$
hyperfine interval and the $D_{21}$ difference can be calculated
using the precise experimental value for $f_{\rm HFS}({1S})$.

A numerical simulation of the two-photon excitation process in the
hydrogen beam shows that the maximum of the excitation probability
distribution for the delayed atoms is shifted in space towards the
first half of the excitation region where the residual magnetic
fields are the smallest. A conservative estimation of a shift
arising from the less shielded paths in the direct vicinity of the
nozzle and the detector gives a value of 0.5(0.5) Hz.

An external electric field $E$ mixes the 2$S_{1/2}$ level with the
adjacent 2$P_{1/2}$ and 2$P_{3/2}$ levels, shifting its energy.
While in first order perturbation theory the $1S$ level is not
shifted in a DC electrical field, the $2S$ $F=0$ and $F=1$ levels
are shifted differently because of their different energy spacing
from the $2P$ levels. According to a calculation of the DC Stark
shift with the hyperfine structure taken into consideration, the
shift of the $2S$ hyperfine interval is equal to $1100 \ E^2$ Hz $
\rm{cm}^2/ \rm{V}^2$. The stray electrical fields within the
excitation region of our setup are estimated to be below 30 mV/cm
\cite{Huber} corresponding to a shift of $-1$ Hz.

The AC Stark shift of a two-photon transition scales inversely to
the energy difference between real levels (in our case $1S$ and
$2S$ levels) and virtual levels \cite{Beausoleil}. The hyperfine
intervals are on the order of one GHz, while virtual energy levels
are about $3/8 \ \rm{Ry}$ away from both $1S$ and $2S$ levels.
Therefore, the differential AC Stark shift of the hyperfine
components in the hydrogen atom is about $10^{-6}$ of the AC Stark
shift of the $ 1S-2S $ transition frequency. The typical AC Stark
shift of the $1S-2S$ transition in our experiment is on the level
of 500 Hz, corresponding to a negligible differential shift of the
$2S$ hyperfine interval assumed that the light intensity is kept
constant.

However, inevitable small fluctuations of the 243 nm light
intensity cause different AC Stark shifts of each hydrogen
spectrum. We have corrected for the intensity fluctuations, using
the experimental value of 2.6 Hz/mW for the $1S-2S$ AC Stark shift
\cite{Niering}, which shifts the final value of the $2S$ interval
by 2 Hz. Besides correction, we have added a conservative 2 Hz
error to the error budget, which may arise from the evaluation of
the light intensity circulating in the enhancement cavity.

Due to the second-order Doppler effect the measured line shape of
the two-photon transition is not symmetric, and the line center is
shifted. For both singlet and triplet transitions, the excited
atoms are from the same atomic beam, and the same velocity class
is selected by the precisely defined delay time. Therefore, the
second-order Doppler effect cancels for the differential
measurement of the $2S$ hyperfine interval. We have evaluated
different velocity classes of hydrogen atoms corresponding to
different delay times and observe essentially no effect on the
evaluated $2S$ hyperfine interval. As an independent test we have
fitted a theoretically simulated lineform for the delay time of
$810\ \mu$s with a Lorentzian function and found that the possible
error of the line center definition is less than 2 Hz. This error
is also added to the error budget.
\begin{figure} [h]
\begin{center}
\includegraphics[width=60mm]{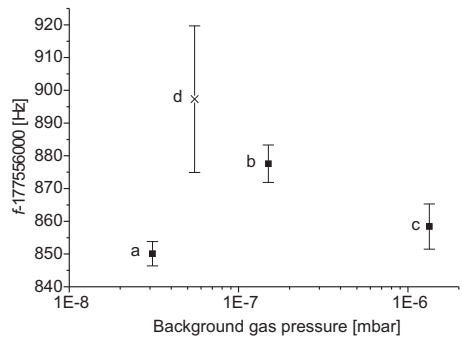}
\caption{Averaged results of measurements of the $2S$ hyperfine
interval at different background gas pressures (logarithmic
pressure scale, error bars give statistical error). The cross
represents a measurement with two times increased hydrogen flow.
a: nine days of measurement, b: 4 days, c: 2 days, d: one
day.}\label{hyspresa}
 \end{center}
\end{figure}

According to \cite{Jamieson}, the interaction cross section for
atomic hydrogen in the $2S$ state is different for triplet and
singlet states, and the pressure shift of the $2S$ hyperfine
interval is comparable to the pressure shift of the $2S$ triplet
level. The previous $2S$ hyperfine interval measurement
\cite{Hessels} indicates for a pressure shift of $-31(24)$
MHz/mbar, which is of the same order of magnitude as the pressure
shift of the $1S (F=1,\ m_F=\pm1)\rightarrow2S (F=1,\ m_F=\pm1)$
transition in hydrogen which is $-8(2)$ MHz/mbar \cite {McIntyre,
Kleppner}.

We have performed four sets of measurements at different
background gas pressures. The data are plotted on
Fig.\ref{hyspresa}, each point representing an averaged result.
Points a, b, c have been measured with approximately the same
hydrogen flow through the nozzle, whereas point d is the result of
a one day measurement with two times increased hydrogen flow.
Within the available range of pressures, we observe no clear
systematic dependence of the 2S hyperfine interval frequency on
the background gas pressure. However, there is some scatter of the
data. The estimated 2S triplet shift due to background gas
pressure is $-8(2)$ MHz/mbar. In the final averaging of the data
points a, b and c, we correct for such a shift, but add a
conservative 10 Hz error to the error budget.

As mentioned above, we have measured for one day in the
differential pumping configuration with the hydrogen flow
increased 2 times (point d on Fig.\ref{hyspresa}) to investigate
the effect of a pressure shift in the hydrogen beam.
Slow atoms interact more frequently with the
rest of the beam than the atoms of average thermal velocity,
therefore the pressure shift should be different for them. We have
evaluated the $2S$ hyperfine intervals for all delay times and
find some non-systematic difference on the level of 5 Hz.
Accounting for the worse statistics of this day of measurement, we
have added an error of 10 Hz for the possible pressure shift in
the hydrogen beam.
\begin{figure} [t]
\begin{center}
\includegraphics[width=50mm]{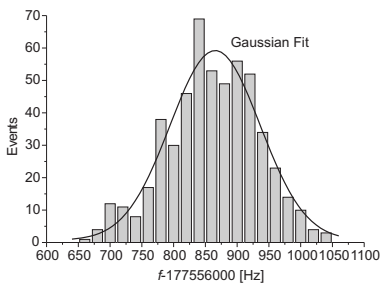}
\caption{Distribution of data points for the $2S$ hyperfine
interval.}\label{hyspresb}
 \end{center}
\end{figure}

One of the main processes causing the data scattering is a
nonlinear drift of the laser frequency on the time scale of 30
min. This process does not cancel in our fit procedure, and has to
be averaged. During the 16 days of measurement the cavity drift
can be considered as random and we expect no systematic shift due
to it. Fig.\ref{hyspresb} represents the distribution of the $2S$
hyperfine interval data without the data of point d. The
distribution is symmetrical and can be approximated with a Gauss
function of 140 Hz width. Statistical averaging yields a value for
$f_{\rm HFS}(2S)$ of $177 \ 556 \ 860(3)$ Hz.

\begin{table} [b]
\caption{Summary of systematic errors and the final result for the
$2S$ hyperfine interval.}
\begin{tabular}{l c c }

  \hline
  \hline
    & Frequency [Hz] & Error [Hz] \\

  \hline
 Averaged interval frequency & 177 556 860 & 3 \\
 Residual magnetic fields & 0.5 & 0.5 \\
 DC Stark shift & -1 & 1 \\
 AC Stark shift & 0 & 2 \\
 Lineshape effects & 0 & 2 \\
 Pressure shift (background gas) & 0 & 10 \\
 Pressure shift in the beam & 0 & 10 \\
 \hline
 final result & 177 556 860 & 15 \\
\hline \hline
\end{tabular}
\label{t1}
\end{table}

\begin{figure} [t]
\begin{center}
\includegraphics[width=60mm]{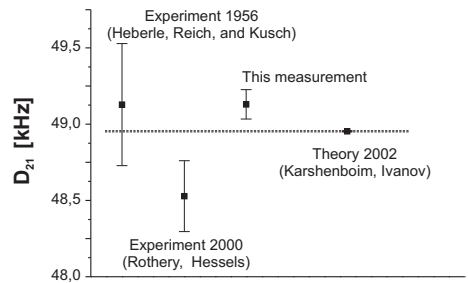}
\caption{$D_{21}$ values corresponding to the several $2S$
hyperfine interval measurements. The dashed lines represent the
 error bar of the theoretical value.}\label{results}
\end{center}
\end{figure}

The effects which contribute errors and shifts to the $2S$
hyperfine interval in our measurement are summarized in Table
\ref{t1}. Fig.\ref{results} represents the $D_{21}$ values
corresponding to the several $2S$ hyperfine interval measurements
in atomic hydrogen as well as the present theoretical value
\cite{Karsh1}. The 15 Hz error, compared to the $2\ 466$ THz
frequency of the $1S-2S$ interval, corresponds to the resolution
of our system on the $6 \times 10^{-15}$ level. The error rivals
the 20 Hz error of the radio-frequency $2S$ hyperfine interval
measurement in deuterium \cite{deuterium}, which can also be
performed optically.

Our current measurement along with other precision experiments on
the hyperfine structure of $1S$ and $2S$ levels in hydrogen and
$^3\rm{He}^+$ ion \cite{Prior} offers a test of QED on level of
accuracy comparable to tests on pure leptonic atoms such a muonium
and positronium  \cite{Karsh1}.

The participation of N.K. in this work was supported by the
Alexander von Humboldt-Stiftung. The work of S.G.K. was supported
in part by RFBR grant \# 03-02-16843. The authors wish to thank A.
Pahl for calculations concerning the DC Stark effect and Eric
Hessels for useful discussions.



\begin{thebibliography}{9.}
%
\bibitem{Kusch} J.W. Heberle, H.A. Reich, and P. Kusch, Phys. Rev. {\bf 101},
612 (1956)
%
\bibitem{Hessels} N.E. Rothery and E.A. Hessels, Phys. Rev. A {\bf 61}, 044501 (2000)
%
\bibitem{Karsh1} S.G. Karshenboim and V.G. Ivanov, Phys. Lett. B {\bf 524}, 259 (2002);
Euro. Phys. J. D {\bf 19}, 13 (2002)
%
\bibitem{Ramsey} N. Ramsey. In: {\em Quantum Electrodynamics}, ed.
by T. Kinoshita (World Scientific, Singapore 1990), p. 673; Hyp.
Interactions {\bf 81}, 97 (1993)
%
\bibitem{Niering} M. Niering {\em et al.}, Phys. Rev. Lett. {\bf 84}, 5496 (2000)
%
\bibitem{Hbook} F. Biraben {\em et al.} In: {\em The Hydrogen Atom. Precision Physics of
Simple Atomic Systems}, ed. by S.G. Karshenboim {et al.}
(Springer, Berlin Heidelberg 2001), p. 17
%
\bibitem{Huber} A. Huber {\em et al.}, Phys. Rev. A {\bf 59}, 1844
(1999)
%
\bibitem{Fischer} M. Fischer {\em et al.},
Can. J. Phys. {\bf{80}}, 1225 (2002)
%
\bibitem{Bethe} H.A. Bethe and E.E. Salpeter, {\em Quantum
Mechanics of One- and Two-Electron Atoms} (Plenum, New York
1977), p. 216
%
\bibitem{Beausoleil} R.G. Beausoleil and T.W. H\"{a}nsch, Phys.
Rev. A {\bf 33}, 1661 (1986)
%
\bibitem {McIntyre} D.H. McIntyre {\em et al.},
Phys. Rev. A {\bf{41}}, 4632 (1990)
%
\bibitem{Kleppner} Th. C. Killian {\em et al.}, 
Phys. Rev. Lett. {\bf{81}}, 3807 (1998)
%
\bibitem{Jamieson} M. Jamieson, A. Dalgarno, and J.M. Doyle, Mol.
Phys. {\bf{87}}, 817 (1996)
%
\bibitem{deuterium} H. A. Reich, J. W. Heberle, and P. Kusch, Phys. Rev. {\bf 104},
1585 (1956).
%
\bibitem{Prior} M. H. Prior, E. C. Wang, Phys. Rev. A {\bf 16}, 6,
(1977)
\end{thebibliography}
\end{document}